\documentclass{nature}
\usepackage{graphicx,caption,subcaption}
\makeatletter
\let\saved@includegraphics\includegraphics
\AtBeginDocument{\let\includegraphics\saved@includegraphics}

\makeatother
\usepackage{amsmath, hyperref}
\usepackage{times}
\usepackage{aas_macros}
\usepackage[super,sort&compress]{natbib}

\begin{document}
\title{Insights into Supermassive Black Hole Mergers from the Gravitational Wave Background }
\author{C.~M.~F. Mingarelli$^1$, L. Blecha$^2$, T. Bogdanovi\'c$^3$, M. Charisi$^{4,5}$, S.~Chen$^{6}$, A. Escala$^{7}$, B.~Goncharov$^{8,9}$, M.~J. Graham$^{10}$, S. Komossa$^{11}$, S.~T.~McWilliams$^{12,13}$, D. A. Schwartz$^{14}$, J. Zrake$^{15}$}

\date{}
\maketitle

\begin{affiliations}
    \item Department of Physics, Yale University, New Haven, 06511, CT, USA
    \item Physics Department, University of Florida, Gainesville, FL 32611, USA
    \item School of Physics and Center for Relativistic Astrophysics, Georgia Institute of Technology, Atlanta, GA 30332, USA
    \item Department of Physics and Astronomy, Washington State University, Pullman, WA 99163, USA
    \item Institute of Astrophysics, FORTH, GR-71110, Heraklion, Greece
    \item Shanghai Astronomical Observatory, Chinese Academy of Sciences, Shanghai 200030, China
    \item  Departamento de Astronom\'ia, Universidad de Chile, Casilla 36-D, Santiago, Chile.
  \item Max Planck Institute for Gravitational Physics (Albert Einstein Institute), 30167 Hannover, Germany
  \item Leibniz Universität Hannover, 30167 Hannover, Germany
  \item Department of Physics, Mathematics and Astronomy, California Institute of Technology, 1200 E California Blvd, Pasadena, CA 91125, USA
  \item Max-Planck-Institut f\"ur Radioastronomie, Auf dem H\"ugel 69, 53121 Bonn, Germany
  \item Department of Physics and Astronomy, West Virginia University, Morgantown, WV 26506, USA 
  \item Center for Gravitational Waves and Cosmology, West Virginia University, Chestnut Ridge Research Building, Morgantown, WV 26505, USA
  \item Smithsonian Astrophysical Observatory, Cambridge, MA 02138, USA
  \item Department of Physics and Astronomy, Clemson University, Clemson, SC 29634, USA
\end{affiliations}

\begin{abstract}
At the Kavli Institute for Theoretical Physics, participants of the rapid response workshop on the gravitational wave background explored discrepancies between experimental results and theoretical models for a background originating  from supermassive black hole binary mergers. Underestimated theoretical and/or  experimental uncertainties are likely to be the explanation. Another key focus was the wide variety of search methods for supermassive black hole binaries, with the conclusion that the most compelling detections would involve systems exhibiting both electromagnetic and gravitational wave signatures.
\end{abstract}

Evidence for the gravitational wave background (GWB) was announced in June 2023 by all major pulsar timing array (PTA) collaborations \cite{NG15-GWB, EPTA-GWB, PPTA-GWB, CPTA-GWB}. There is little doubt that the signal originates from low-frequency gravitational waves in the nanoHertz (nHz) regime, however tighter constraints on the GWB's strain spectrum, or signs of discreteness therein\cite{Discreteness2024}, are needed to confirm that the signal originates from supermassive black hole binary (SMBHB) mergers. Furthermore, the amplitude of the background is $2-4.5~\sigma$ above predictions of most models \citep{satopolito2024}. There may be multiple reasons for this, including but not limited to underestimated black hole masses, underestimated errors in the galaxy stellar mass functions used in theoretical predictions of the background, and mis-modeled pulsar noise, e.g.~\cite{larsen24}. We explored these topics, as well as the multimessenger signals for identifying individual SMBHB systems, at the KITP rapid response workshop, 
\href{https://www.kitp.ucsb.edu/activities/stronggwb-m24}{Gravitational Wave Background Found in Pulsar Timing Arrays: Implications for Merging Supermassive Black Holes}, from November 12 - 22, 2024. Due to the Rapid Response nature of this workshop, attendance was limited to 20 participants. The first week of the workshop focused on the GWB, while the second was centered on discussions surrounding electromagnetic counterparts to SMBHBs. 

The workshop started with a talk by Siyuan Chen, who provided an overview of previous GWB upper limits, leading to evidence of the background in NANOGrav, EPTA and InPTA, PPTA, and CPTA data. Chiara Mingarelli then delivered a UCSB colloquium on PTAs, which included all KITP attendees and members of the Department. Our next gathering was in the form of a panel discussion led by Gabriela Sato-Polito and Luke Kelley, focusing on the underlying SMBHB population potentially sourcing the GWB. We learned about the new Velocity Dispersion Function (VDF) GWB model~\cite{satopolito2024}. The VDF model connects the velocity dispersions of galaxies to the SMBHB population using the M--$\sigma$ relation, providing an observationally-based framework to predict the GWB. The VDF model is a classed as a ``Major Merger'' model, which relies galaxy mergers as the primary formation channel for SMBHBs, linking galaxy mergers with SMBHB mergers via a time delay caused by dynamical processes such as dynamical friction, stellar hardening, and gas-driven inspirals, to evolve SMBHBs into the nHz regime, generating a GWB. In addition to the VDF model, we examined Ref~\cite{2014ApJ...789..156M}, another Major Merger model, which gave an effective theoretical upper limit to how loud a GWB signal can be, and agrees well with the amplitude observed by PTAs.

The last talk of the week was given by Boris Goncharov and Mingarelli on pulsar noise. They presented recent results showing how advances in pulsar noise models can improve GWB characterization, and enhance the confidence of GWB detection. Goncharov shared recent results from the EPTA \cite{GoncharovSardana2024b} where improved noise models reduced the GWB amplitude and aligned its spectral index more closely with $\gamma=13/3$. This brings the GWB measurement more in line with a SMBHB origin, Figure 1. Mingarelli presented complementary results from NANOGrav, focusing on J1713+0747 \cite{larsen24}.

In the second week, we focused on the variety of possible electromagnetic counterparts to sub-parsec SMBHBs. Maria Charisi gave a detailed overview of periodic photometric variability of Seyfert galaxies and quasars~\cite{Dorazio2023, Komossa2024}, which is one of the main expected electromagnetic signals.
We discussed at length how many periods we would need to see in an Active Galactic Nucleus (AGN) light curve before we would believe it is really periodic, given the presence of red noise~\cite{Witt2022}. The consensus was that we need to see at least 5 periods, depending on the data quality and statistical confidence of the claim.
Furthermore, Charisi reviewed evidence that light curves of (single) active AGN are more complicated than damped random walks and highlighted the need for improved noise modeling when searching light curves for the (faint) periodic signal due to orbital motion in a binary system~\cite{Robnik2024}. We also discussed the synergies between electromagnetic time-domain surveys and PTAs~\cite{Charisi2022}, as well as the challenges in identifying the host galaxy of an individual SMBHB in the large PTA localization area of an all-sky search~\cite{2024ApJ...976..129P}.
In the afternoon, Kelley delivered a Chalk Talk for all KITP participants. 

In our second meeting of the second week, Haowen Zhang presented the \textsc{Trinity} model, which reconstructs galaxy and black hole assembly histories and predicts SMBHB populations and their gravitational wave-emission by modeling SMBHB evolution with post-processing. The post-processing incorporates dynamical friction and gravitational wave-driven evolution, and reproduces PTA observations.
We then had a panel discussion on electromagnetic counterparts, with panelists Tamara Bogdanović, Laura Blecha, Jonathan Zrake, Matthew Graham, and Stefanie Komossa.
Electromagnetic signals include periodic flux variability in AGN caused, for instance, by Doppler boosting (particularly important for unequal mass binaries), Doppler shifts of broad emission lines, precessing radio jets, characteristic broad-band spectral energy distributions, and variable X-ray line emission~\cite{Dorazio2023}. 
The panel addressed the challenges of associating these signals uniquely with SMBHBs, as opposed to physical processes around single SMBHs which can potentially produce similar signatures, and we emphasized the need for combining the gravitational wave and electromagnetic signals for most reliable binary identification.
We also re-discussed red noise in the AGN light curves leading to false positives, and what one would need to see in order to believe a multimessenger SMBHB detection. The answer to this question remains open, however many agreed that seeing the gravitational wave frequency evolve over the PTA observation, if possible, would be very compelling.

Finally, Zrake led a
lively and insightful discussion on dynamics of SMBHBs in circumbinary disks. Circumbinary disks form eccentric cavities, with gas streams creating shock-heated mini-disks around each SMBH. Preferential accretion onto the lower-mass black holes drives the binary's mass ratio to equalize. He discussed why mass and angular momentum exchange with a circumbinary disk is likely the main driver of SMBHB orbital hardening at frequencies below 1 nHz, whereas in the PTA band, SMBHB inspirals are expected to be dominated by gravitational wave losses. He also described the viscous decoupling process, which leads to a reduction of gas delivery to the black holes in the final stages of the gravitational wave-driven inspiral --- hours to centuries, depending on black hole mass and disk properties. Zrake highlighted that the timescale for viscous decoupling is very sensitive to the effective viscosity of AGN disks, and thus remains a significant source of uncertainty in the prediction of electromagnetic signals from inspiraling SMBHBs.

The meeting concluded with a forward-looking discussion on how emerging facilities and improved PTA techniques will shape the field. The Deep Synoptic Array-2000 (DSA-2000) and the Square Kilometer Array will enhance the detection of radio counterparts, and find more pulsars, while the Rubin Observatory and its Legacy Survey of Space and Time (LSST) will lead optical variability studies. Expanding PTA networks, incorporating more millisecond pulsars, and extending timing baselines are critical steps for improving PTA sensitivity. 

Leaving the meeting, participants noted the great efficiency of the KITP workshops in bringing together colleagues from different backgrounds --- in this instance the PTA and the SMBH communities. They commented on how this triggers excellent discussions, the development of new ideas and collaborations, and a variety future projects.

\begin{figure*}
    \centering
    \begin{subfigure}[b]{0.61\textwidth}
        \includegraphics[width=\textwidth]{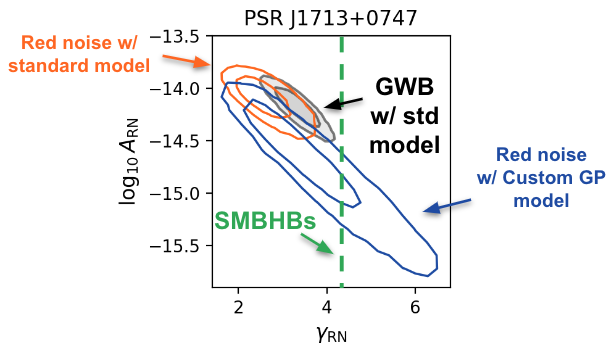}
        \caption{One NANOGrav pulsar} 
        \label{fig:pulsar-noise:nanograv}
    \end{subfigure}
    \begin{subfigure}[b]{0.38\textwidth}
    \includegraphics[trim={0 0 3.8cm 3.7cm},clip,width=\textwidth]{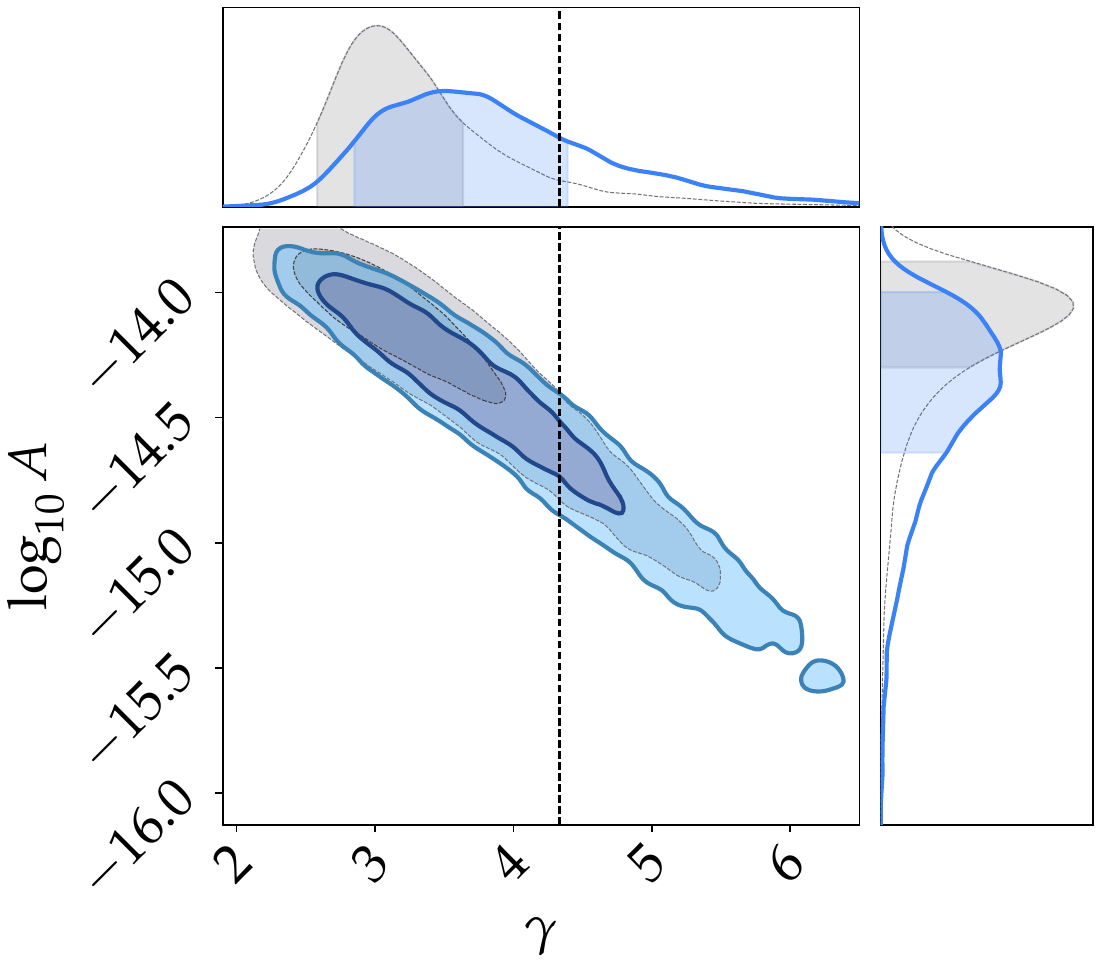}
        \caption{All EPTA pulsars}
        \label{fig:pulsar-noise:epta}
    \end{subfigure}
    \caption{Pulsar noise can affect the GWB amplitude and spectral index measurements. \textit{Left}: Chromatic noise is more successfully mitigated using Gaussian Processes, leading to lower amplitude red noise (blue) than the standard analysis (orange) for PSR J1713+0747~\citep{larsen24}. Importantly, J1713+0747's new red noise is more consistent with a GWB from SMBHBs, (green line). \textit{Right}: The impact of an improved noise model on the inferred strain amplitude and spectral index of the GWB measured by the EPTA~\citep{GoncharovSardana2024b}.}
    \label{fig:pulsar-noise}
\end{figure*}

 \section*{Acknowledgments}
 We thank Lars Bildsten, David Kaczorowski, and Bibi Rojas for making our stay at KITP enjoyable and productive. We also thank Bjorn Larsen for providing the left panel for Figure 1. Our work was supported in part by NSF grants: PHY-2309135 to the Kavli Institute for Theoretical Physics (KITP), NSF PHY-2020265, AST-2414468, NSF AST-2307278, and NASA grants 80NSSC22K0748, 80-NSSC-24K0440. This work was also supported by the Flatiron Institute, part of the Simons Foundation, the Research Corporation for Science Advancement from award CS-SEED-2023-008, ANID FB-210003, and by the European Union (ERC-MMMonsters-101117624).
\bibliographystyle{naturemag}

\end{document}